# Anion Exchange in II-VI Semiconducting Nanostructures via Atomic Templating

*Rahul Agarwal[1], Nadia M. Krook[1], Ming-Liang Ren[1], Liang Z. Tan[2], Wenjing Liu[1], Andrew M. Rappe[2], Ritesh Agarwal[1]\**

[1]Department of Materials Science and Engineering, University of Pennsylvania, Philadelphia, PA 19104, USA

[2] Department of Chemistry, University of Pennsylvania, Philadelphia, PA 19104-6323, USA

\* To whom correspondence should be addressed. E-mail: riteshag@seas.upenn.edu

**Controlled chemical transformation of nanostructures is a promising technique to obtain precisely designed novel materials which are difficult to synthesize otherwise. We report high-temperature vapor phase anion exchange reactions to chemically transform II-VI semiconductor nanostructures (100-300 nm length scale) while retaining the single crystallinity, crystal structure, morphology and even defect distribution of the parent material via atomic templating. The concept of atomic templating is employed to obtain kinetically controlled, thermodynamically metastable structural phases such as zincblende CdSe and CdS from zincblende CdTe upon complete chemical replacement of Te with Se or S. The underlying transformation mechanisms are explained through first-principles density functional theory calculations. Atomic templating is a unique path to independently tune materials' phase and composition at the nanoscale allowing synthesis of novel materials.**

Ion-exchange reactions, i.e., selective chemical substitution of ions in ionic materials, has been utilized as an effective route for chemical transformation of nanostructures to design novel materials with unique physical, chemical and structural properties and also to understand the distinctive transformation mechanisms at the nanoscale[1-14]. However, most ion exchange studies in nanostructures have focused on cation exchange in sub-10 nm crystals via solution phase chemistries; these studies have provided important insights about cation replacement reactions at these small length scales, but have limited the scope in understanding chemical transformations in intermediate length scale nanostructures (100 – 300 nm) such as nanowires (NWs) and nanobelts (NBs), with critical dimensions in this size scale. We previously reported cation exchange in CdS NWs with diameters in the 10-300 nm range with Zn, via high temperature vapor phase reactions. This enables kinetically-controlled and size-dependent chemical transformation products ranging from alloyed NWs to metal-semiconductor heterostructures, underscoring the importance of chemical transformations in nanostructures for the synthesis of novel materials[15]. Anion exchange reactions, on the other hand, are kinetically quite challenging in comparison to cation exchange, because anions are usually much bigger than their cationic counterparts in a compound. Anions form the structural framework in crystals and are therefore more difficult to dislodge without upsetting the structural integrity of the material[16]. Anions possess sluggish diffusivity owing to their large size which renders their replacement with another anion and subsequent ejection from the material kinetically unfavorable. Therefore, there are limited reported studies on anion exchange reactions in nanostructures, and much of the work has been accomplished through solution phase ion exchange routes[17-24]. The thermal energy of solution phase chemical reactions is only capable of supporting diffusion over very short distances (<50 nm) in typical laboratory experiment time frames and hence this is not the preferred route to effect chemical transformations in larger (100-300 nm) nanostructures. Therefore, one needs to develop alternate routes to replace anions in nanostructures over longer length scales, which would require higher temperatures and better control over reactant delivery to precisely control the rate of exchange reactions. Proper control over reaction rates via temperature and reactant control is essential because very rapid exchange processes would destroy the structural integrity of the sample, and very low reaction rates (typically at lower temperatures) may not be adequate to replace anions deeper in the structure. Here, via high temperature vapor phase chemical reactions, we report anion exchange reactions in II-VI semiconducting NWs and NBs, structures that are excellent systems to access a variety of lengthscales and morphologies along with significant ionicity to study their transformation properties. II-VI nanostructures are also important due to their interesting physical properties and potential nanoelectronic and nanophotonic applications[25-28]. In our studies, we achieve complete anion exchange in a variety of nanostructures of different sizes while retaining the atomic template of the parent compound in the form of its crystal structure, defect distribution and morphology. These observations can enable precise engineering of well-known semiconductor compositions in new and previously unobserved kinetically controlled phases, which are difficult to obtain otherwise, leading to synthesis of new materials with novel properties.

To explore the potential of anion exchange for the chemical transformation of nanostructures and to study the accompanying evolution in the crystal structure and physical properties, single-crystalline wurtzite (WZ) CdS nanostructures were synthesized via the vapor-liquid-solid (VLS) method in a tube furnace set-up[27], and anion exchange was performed on as-grown CdS nanostructures with Se vapor (see supplementary information for experimental details). The experimental conditions were optimized after extensive trials involving variation of precursor type, quantity, reaction temperature, and time, and it was observed that for a reaction time of approximately one hour, the temperature was the most important factor which determined the final product. Overheating (> 625 °C) caused excessive sublimation, and underheating (< 575 °C) leads to negligible chemical transformation. Anion exchange reaction at 600°C for one hour was observed to be optimal to obtain complete chemical transformation of CdS into CdSe while retaining the single-crystalline WZ structure, morphology and the growth orientation of the parent nanostructure (Figure 1 A-E). The results were reproduced for a wide range of sizes and morphologies of WZ CdS ranging from sub-100 nm diameter NWs to 500 nm wide NBs. The observed surface roughness of the nanostructures can be attributed to ≈4% lattice expansion upon transformation from WZ CdS to WZ CdSe and some material deposition on the surface accompanying high temperature anion exchange reaction. Raman spectroscopy of a completely transformed CdSe nanostructure shows that longitudinal optical (LO) phonon peaks corresponding to pure CdSe[29] were present while those corresponding to pure CdS[30] or $CdS_xSe_{1-x}$ alloys[31] were absent, confirming the presence of only Cd-Se bonds in the transformed product (Figure 1F). Photoluminescence (PL) spectroscopy of a completely transformed CdSe NW indicates band edge emission corresponding to pure CdSe[32] (≈716 nm) (Figure 1 G), further confirming that the band gap of the anion exchanged material matched WZ CdSe. Achieving complete chemical anion exchange in large nanostructures while retaining the structural integrity of the parent material is a surprising result, which we attribute to controlled high temperature vapor phase reaction that enabled large-scale atomic diffusion and substitution reaction, while maintaining relatively slow reaction rates so as not to collapse the structure. It is important to note that while the anions, which are the structural pillars of the lattice, were being replaced, the cations remained in their positions, thereby retaining the single-crystalline structure of the parent compound. Although, chemical transformation of CdS into CdSe is thermodynamically unfavorable[33] (see supplementary information), the high temperature and the vapor phase Se reactant succeeded in kinetically driving forward this reaction. The exact mechanism of this chemical substitution process will be described in detail later.

Motivated by the evidence that the anion lattice framework can be preserved during the anion exchange reaction in WZ CdS, the concept was further explored in other phases of II-VI semiconductors with the idea of independently engineering the crystal structure and chemical composition of the material. For example, CdTe is thermodynamically stable in the zincblende (ZB) crystal structure, whereas CdS is stable in the WZ crystal structure at room temperature and atmospheric pressure[34,35]. However, CdSe crystal structure stability at ambient conditions is size-dependent; typically, <10 nm diameter CdSe nanocrystals can exist in the ZB phase, whereas larger crystals stabilize in the WZ phase[36-45]. Therefore, it is challenging to obtain larger (>100 nm) crystals (NWs and NBs) of CdSe in zincblende (ZB) phase via conventional synthesis techniques. Since the anion framework can be maintained during the replacement reactions, we performed experiments in order to obtain metastable crystal structures (e.g., CdSe in ZB phase). Anion exchange reaction was performed in ZB CdTe NWs and NBs of different sizes (>100 nm diameter, thickness and widths) with vapor phase transport of Se vapor at 450°C. Transformation of CdSe from CdTe requires lower temperature (450°C) in comparison to CdSe from CdS (600°C) owing to lower bond energy of Cd-Te than Cd-S, as also reflected by their respective melting temperatures[34,35]. In order to understand the evolution of crystal structure upon chemical transformation, the same nanostructure assembled on e-beam transparent $SiN_x$ membrane was identified and characterized in detail via TEM studies before and after the replacement reaction. In agreement with our expectations, we obtained CdSe nanostructures in the ZB phase upon anion exchange of Te by Se in ZB CdTe, and the resulting materials retained the morphology, single crystallinity and most interestingly, the ZB crystal structure of parent CdTe upon complete chemical transformation (Figure 2 A-C). The lattice constant of ZB CdSe was measured to be ≈6.1 Å, which closely matches the calculated value for ZB CdSe, assuming that ZB and WZ have similar Cd-Se bond distances of ≈2.6 Å (see supplementary information for calculations). Raman and PL spectroscopy also confirmed complete chemical transformation by displaying LO phonon and band-edge emission PL peaks corresponding to pure CdSe[46,47] (Figure 2 D-E). Single crystalline ZB CdSe formation from ZB CdTe indicates that high temperature vapor phase anion exchange reaction can preserve the atomic template of the parent nanostructure.

If anion exchange is truly atomic template preserving, it must also retain the nature and location of extended lattice defects in a nanostructure upon chemical transformation, assuming that the reaction temperature does not exceed the defect annealing temperature itself. To validate this hypothesis, ZB CdTe nanostructures synthesized via the VLS mechanism[48], which naturally possess planar growth defects such as twin boundaries[49] (see supplementary information for characterization details), were chosen for anion exchange reaction with Se (reaction temperature, 450 ºC). The same nanostructure was characterized before and after the anion exchange reaction to study the evolution of defects and their distribution upon chemical transformation (Figure 3 A). Supporting our hypothesis of atomic templating, the resulting NW not only retained the single-crystalline ZB crystal structure but also the periodic twin boundaries of the parent material upon undergoing complete chemical transformation into CdSe (Figure 3 B-C). These results in addition to providing important details about the chemical transformation mechanism (discussed later), show the potential for independently engineering both the chemical composition and structural phase of materials at the nanoscale while preserving the defect distribution of the parent material.

As mentioned earlier, CdS is thermodynamically stable in the WZ crystal structure at room temperature and atmospheric pressure, and it is challenging to synthesize larger nanostructures (>100 nm) of CdS in the ZB phase[42-45]. Therefore, to obtain large ZB CdS nanostructures, we performed anion exchange reaction at 450 ºC in ZB CdTe nanostructures with S vapor. Upon performing the reaction, we once again observed complete chemical transformation into single-crystalline ZB CdS while retaining the twin boundaries of the parent CdTe NBs (Figure 4). The lattice constant of ZB CdS was measured to be ≈5.8 Å, which closely matches the calculated value for ZB CdS structure, assuming that ZB and WZ have similar Cd-S bond distance of ≈2.5 Å (see supplementary information for calculations). Successful chemical transformation of ZB CdTe into ZB CdS while retaining the planar defects is another proof of the lattice preserving nature of the anion exchange reaction a the nanoscale. The ability to synthesize large nanostructures of CdS and CdSe in ZB phase highlights the role of anion exchange based chemical transformation technique in phase engineering of metastable structures which are challenging to obtain otherwise.

To understand the mechanism behind the observed lattice-framework-preserving chemical transformation, a series of control experiments were designed to study anion exchange in the WZ CdS model system by introducing Se. The reaction was kinetically controlled by lowering the reaction temperature, thereby decelerating the transformation kinetics while keeping all other experimental conditions the same, to obtain insights into the step by step structural and chemical evolution of the nanostructures. Upon performing anion exchange at 300ºC (lowered from 600ºC) in a ≈200 nm diameter WZ CdS NW for one hour, we observed that the NW possesses a rough morphology with periodic protrusions on the surface. Selected area electron diffraction (SAED) patterns of the NW reveal single-crystalline WZ structure with split diffraction spots corresponding to pure WZ CdS and WZ CdSe (Figure 5 A). An EDS line scan along the radial direction indicates that the NW possesses a compositional gradient, with Se-rich phase on the surface and S-rich phase toward the core, indicating incomplete anion exchange reaction (Figure 5 B). However, upon performing anion exchange at 450ºC on another ≈200 nm diameter WZ CdS NW for one hour, we again observe rough surface morphology (smoother than the experiment at 300ºC but still much coarser than the reaction at 600ºC) but we find a uniform lattice parameter (no observable diffraction spot splitting) throughout the material (Figure 5 C). EDS line scan along the radial direction indicates a compositional gradient. However, the penetration depth of Se toward the core is higher than was obtained via anion exchange reaction performed at 300ºC (Figure 5 D). Raman spectroscopy of the NW indicates LO phonon peaks corresponding to both Cd-S and Cd-Se bonds in a $CdS_xSe_{1-x}$ type alloy in addition to a sum frequency peak (≈480 cm$^{-1}$) (Figure 5 E) that corresponds to a two-mode behavior from the $CdS_xSe_{1-x}$ alloy[31].

Based on the results obtained from partial and complete chemical transformation in WZ CdS with Se, we propose the following mechanism for the observed anion exchange reaction. The reaction temperature influences the rate of the reactant species generated (Se vapor), the kinetic driving force for the anion exchange to proceed and the diffusivity of atoms in the nanostructure, all of which determine the chemical composition and composition distribution of the final product. At a particular reaction temperature, CdS nanostructures (Figure S4A) will be exposed to a constant flux of Se vapor which will react with CdS at the surface creating domains of Se-rich $CdS_xSe_{1-x}$ alloy through anion exchange, and the excess vapor will condense as amorphous selenium (a-Se) on top creating protrusions (Figure S4B). Besides anion exchange, incoming Se atoms can occupy the S vacancies sites in CdS[50,51], which is a thermodynamically favorable process. Our first-principles DFT calculations (see supplementary information) show that the Gibbs free energy for the reaction involving one Se atom in vapor occupying a single S vacancy in WZ CdS is $\Delta G$ = -3.01 eV/vacancy at 600°C, $P_{Se}$ = 100 Torr.

Unlike the thermodynamically highly favorable vacancy occupation reaction, the anion substitution reaction depends strongly on the reaction conditions. Our calculations show that the excess Se source vapor and high temperature play a major role in driving this reaction. We find that anion substitution proceeds when S is at a much lower partial pressure than Se:

WZ CdS + Se (gas) → WZ CdSe + S (gas)

$\Delta G_{rxn}$ = -0.11 eV/unit cell at 600°C, $P_{Se}$ = 100 Torr and $P_S$ = 10$^{-4}$ Torr

In our experiment, the carrier gas (Argon flowing at 15 SCCM) is capable of instantaneously transporting away rejected S vapor, which is a byproduct of the forward anion exchange reaction, acting as an efficient sink for S. This helps in pushing the reaction forward and minimizing the reverse anion exchange reaction (CdSe transforming into CdS) reaction. On the other hand, at equal partial pressures of Se and S vapor, our calculations predict that complete transformation of WZ CdS into WZ CdSe at our reaction temperature (600°C) via anion exchange is thermodynamically unfavorable; $\Delta G_{rxn}$ = 0.932 eV/unit cell at 600°C, $P_{Se}$ = $P_S$ = 100 Torr. This shows that lowering S vapor partial pressure as compared to Se vapor partial pressure reduces the Gibbs free energy of the products, making the forward reaction thermodynamically favorable by lowering $\Delta G_{rxn}$ from 0.93 eV/unit cell to -0.11 eV/ unit cell at 600°C and the backwards reaction kinetically unfavorable by increasing the activation barrier as a result of reducing Gibbs free energy of products. Upon anion exchange, the initial $CdS_xSe_{1-x}$ domain size will be directly proportional and domain spacing will be inversely proportional to the incoming Se vapor flux, which is dictated by the reaction temperature. Domain growth of $CdS_xSe_{1-x}$ will involve outward diffusion of $Se^{2-}$ to the $S^{2-}$ rich regions while fresh $Se^{2-}$ is introduced into the system through constant anion exchange (Figure S4 C). In addition, surface diffusion, which is faster than bulk diffusion, will lead to preferential chemical transformation at the surface before the core of the nanostructure, thereby creating core-shell type morphology instead of an axial heterostructure (Figure S4 D). Since the source of Se vapor can be assumed to be infinite with respect to the limited nanostructure volume available, the anion exchange reaction and $CdS_xSe_{1-x}$ domain growth will continue on the surface until heating is stopped (Figure S4 E). Therefore, at the lowest reaction temperature (300°C) for one hour, pure CdS at the core is preserved and is surrounded by a shell of $CdS_xSe_{1-x}$ alloy which explains the diffraction spot splitting in SAED pattern (Figure 5 A inset). The rough surface morphology is due to $Cd^{2+}$ diffusing from nearby regions into the accumulated a-Se protrusions, thus creating Se-rich $CdS_xSe_{1-x}$ in the outermost shell. At an intermediate reaction temperature (450°C) for an hour, the entire nanostructure transforms into $CdS_xSe_{1-x}$ alloy with a radial composition gradient which allows the system to relax to an intermediate lattice parameter between pure CdS and CdSe. The surface roughness is relatively lower, since the a-Se condensation rate will be smaller at higher temperatures, and the deposited a-Se will be consumed at a faster rate through anion exchange. At the highest reaction temperature (600°C), the reaction front proceeds inwards all the way to the core, completely transforming CdS into CdSe with the lowest surface roughness, due to negligible a-Se deposition on the surface. The most important factor responsible for the observed surface roughness at 600°C is lattice expansion by ≈4% as CdS converts into CdSe and also material sublimation due to large strain and high temperature of the reaction. Since the fully transformed nanostructures show no signature of any residual sulfur, it is believed to diffuse outward to the surface and then sublimate. At reaction temperatures below 300°C, no anion exchange is observed for similar reaction duration (one hour) since there is not enough kinetic drive to push forward both the chemical substitution reaction and atomic diffusion. At reaction temperatures higher than 600°C, the nanostructures sublimate, possibly owing to very fast chemical substitution reaction, thereby causing large scale bond breaking and thereby destabilizing the anion lattice framework. These experiments demonstrate the utility of vapor phase anion exchange reactions in chemically transforming large nanostructures to new single composition or compositionally graded nanomaterials while preserving the lattice framework of the parent material. This enables the synthesis of metastable phases of materials at the nanoscale while preserving the defect density of the parent material.

To summarize, we have demonstrated vapor phase anion exchange reactions as a novel chemical transformation route to independently control material's structural phase and composition, thereby allowing us to synthesize well-known compounds in metastable stable crystal phases with precise defect distribution which is retained from the parent nanostructure (Figure 6). Our work exhibits controlled chemical reactions at the nanoscale while preserving the structural integrity of the material in spite of substituting anions which are considered structural pillar in a crystal framework. This method, which we call "atomic templating", is a promising tool that can be extended beyond II-VI semiconductors to engineer more exotic materials at the nanoscale for a variety of electronic, optical, and energy harvesting and conversion applications.

**Acknowledgements**: R.A. and L.Z.T. were supported by NSF grant DMR-1120901. A.M.R acknowledges support from the US Department of Energy under grant DE-FG02-07ER46431. Transmission electron microscopy experiments were performed at the Singh Center for Nanotechnology at the University of Pennsylvania. Raman spectroscopy was performed with the support of the Nano/Bio Interface Center through the National Science Foundation Major Research Instrumentation Grant DMR-0923245.

**Author contributions:** R.A. and R.A.* conceived and designed the experiments. R.A., N.M.K. and W.L. optimized the growth conditions and synthesized the CdS and CdTe nanostructures. R.A. performed the anion exchange reaction and the chemical and structural characterization before and after it. M-L.R. performed photoluminescence experiments. L.Z.T and A.M.R performed first principles density functional theory calculations. R.A. and R.A.* interpreted and analyzed the data. R.A., L.Z.T., A.M.R., and R.A.* co-wrote the manuscript.

*denotes corresponding author

**Associated Content**

TEM, Raman Spectroscopy, and Photoluminescence Spectroscopy experimental details, growth and characterization details of CdTe nanobelts, anion exchange experimental details, ZB lattice constant calculations for CdSe and CdS, Figures S1-S4, first-principles calculations using density functional theory.

**Figure Captions**

**Figure 1: Characterization of nanostructures after undergoing complete anion exchange from WZ CdS into WZ CdSe.** A) Bright field-scanning TEM (BF-STEM) micrograph and EDS map scan of a ≈200 nm diameter NW after complete chemical transformation from CdS into CdSe. B) BF-TEM micrograph of the same NW confirming morphological integrity along the entire length after anion exchange. Inset: SAED pattern corresponding to single-crystalline WZ CdSe. C) BF-TEM micrograph of a NB after undergoing complete chemical transformation. Inset: SAED pattern corresponding to single-crystalline WZ CdSe. D) BF-STEM micrograph and EDS map of the same CdSe NB showing uniform chemical transformation. E) EDS point scan from the same NB confirming complete chemical transformation from CdS to CdSe with no residual S. F) Raman spectrum of a completely transformed CdSe nanostructure showing LO phonon peaks corresponding to Cd-Se bond. The large background is due to the photoluminescence (PL) emission due to 532 nm laser excitation. G) PL spectrum of a completely transformed CdSe nanostructure showing band edge emission corresponding to pure CdSe.

**Figure 2: Characterization of NW after undergoing complete anion exchange from ZB CdTe into ZB CdSe.** A) Bright-field TEM micrograph of a ZB CdTe NW before anion exchange. Inset: SAED pattern confirming the ZB crystal structure and <111> growth axis. B) Bright-field TEM micrograph of the same CdTe NW after complete anion exchange into CdSe. Inset: SAED pattern confirming retention of the ZB crystal structure and <111> growth axis with lattice parameters matching pure ZB CdSe. C) EDS point scan of the same NW, confirming

complete chemical transformation into CdSe with no residual Te. D) The Raman spectrum also confirms complete chemical transformation with the presence of only Cd-Se LO phonon peaks. E) PL spectrum displaying band-edge emission corresponding to pure CdSe.

**Figure 3: Characterization of a twinned NW after undergoing complete anion exchange from ZB CdTe into ZB CdSe.** A) Dark-field TEM micrograph of a periodically twinned CdTe NW before anion exchange. Upper Left Inset: SAED pattern confirming the ZB crystal structure, twinning and <111> growth axis. Bottom Right Inset: Bright-field TEM micrograph of the same NW. B) Dark-field TEM micrograph of the same NW after complete anion exchange into CdSe. Upper Right Inset: SAED pattern confirming retention of the ZB crystal structure, twinning and <111> growth axis with lattice parameters matching pure ZB CdSe. C) EDS point scan of the same NW confirming complete chemical transformation into CdSe with no residual Te.

**Figure 4: Characterization of a NB after undergoing complete anion exchange from ZB CdTe into ZB CdS.** A) Bright-field TEM micrograph of a ZB CdS NB after undergoing complete chemical transformation from ZB CdTe. Inset: SAED pattern of the NB confirming retention of ZB crystal structure, twinning and <112> growth axis. B) EDS map scan of the NB showing uniform chemical transformation into CdS. C) EDS point scan of the NB confirming complete chemical transformation. D) Raman spectrum displaying LO phonon peaks corresponding only to Cd-S bonds, thereby confirming complete chemical transformation. E) PL spectrum displaying band edge emission corresponding to pure CdS.

**Figure 5: Characterization of nanostructures after undergoing partial anion exchange from CdS to CdS$_x$Se$_{1-x}$.** A) Bright-field TEM micrograph of CdS$_x$Se$_{1-x}$ NW after anion exchange at 300$^0$C. Inset: SAED pattern of the NW confirming single-crystalline WZ structure with split and elongated spots (red circles). B) EDS line scan of the same NW confirming the core-shell morphology of the alloyed NW (along the dotted line in A). C) Bright-field TEM micrograph of CdS$_x$Se$_{1-x}$ NW after anion exchange at 450$^0$C. Inset: SAED pattern of the NW, confirming single-crystalline WZ structure of CdS$_x$Se$_{1-x}$. D) EDS line scan of the same NW confirming the core-shell morphology of the alloyed NW (taken along the dotted line in C). E) Raman spectrum of an alloyed CdS$_x$Se$_{1-x}$ NW showing LO phonon peaks corresponding to both Cd-S and Cd-Se bonds.

**Figure 6:** Schematic representation of the concept of atomic templating where, for example, allotropes (WZ and ZB) of a compound (CdSe) can be produced by performing anion exchange in starting compounds with different crystal structures (WZ CdS versus ZB CdTe).

# Figure 1

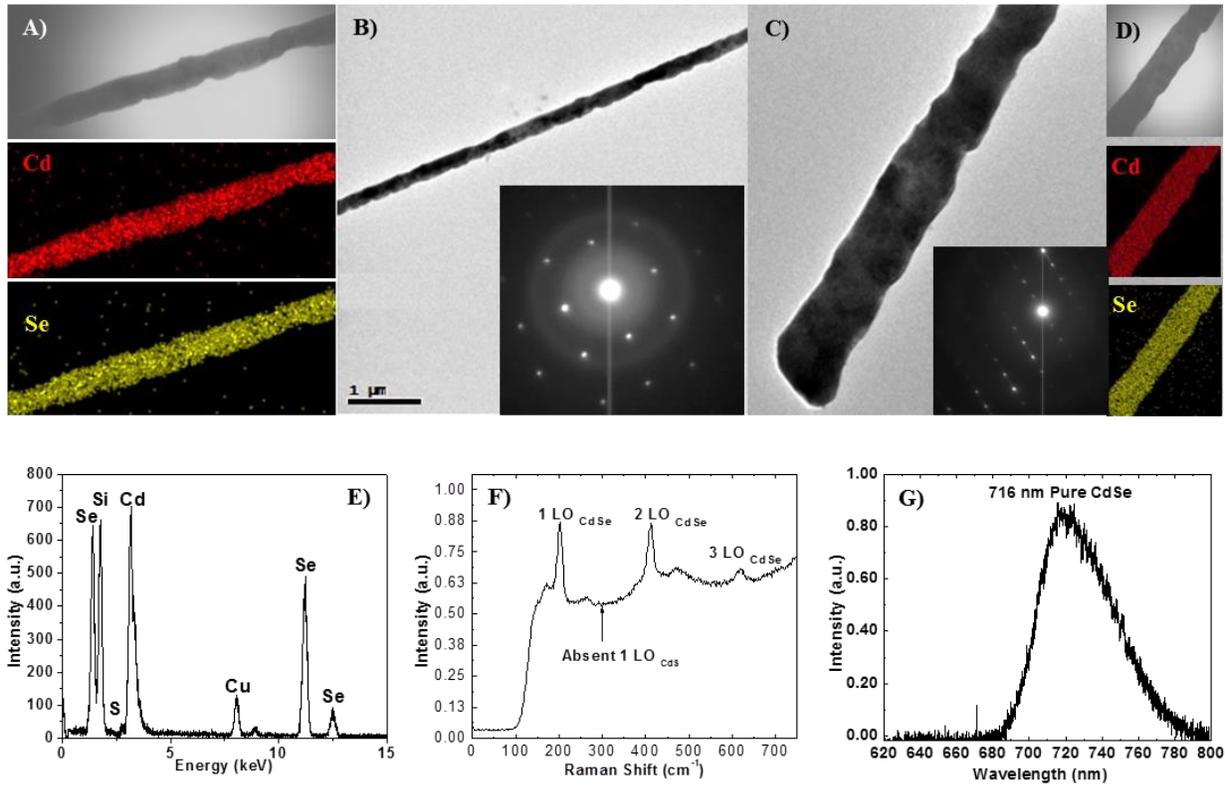

# Figure 2

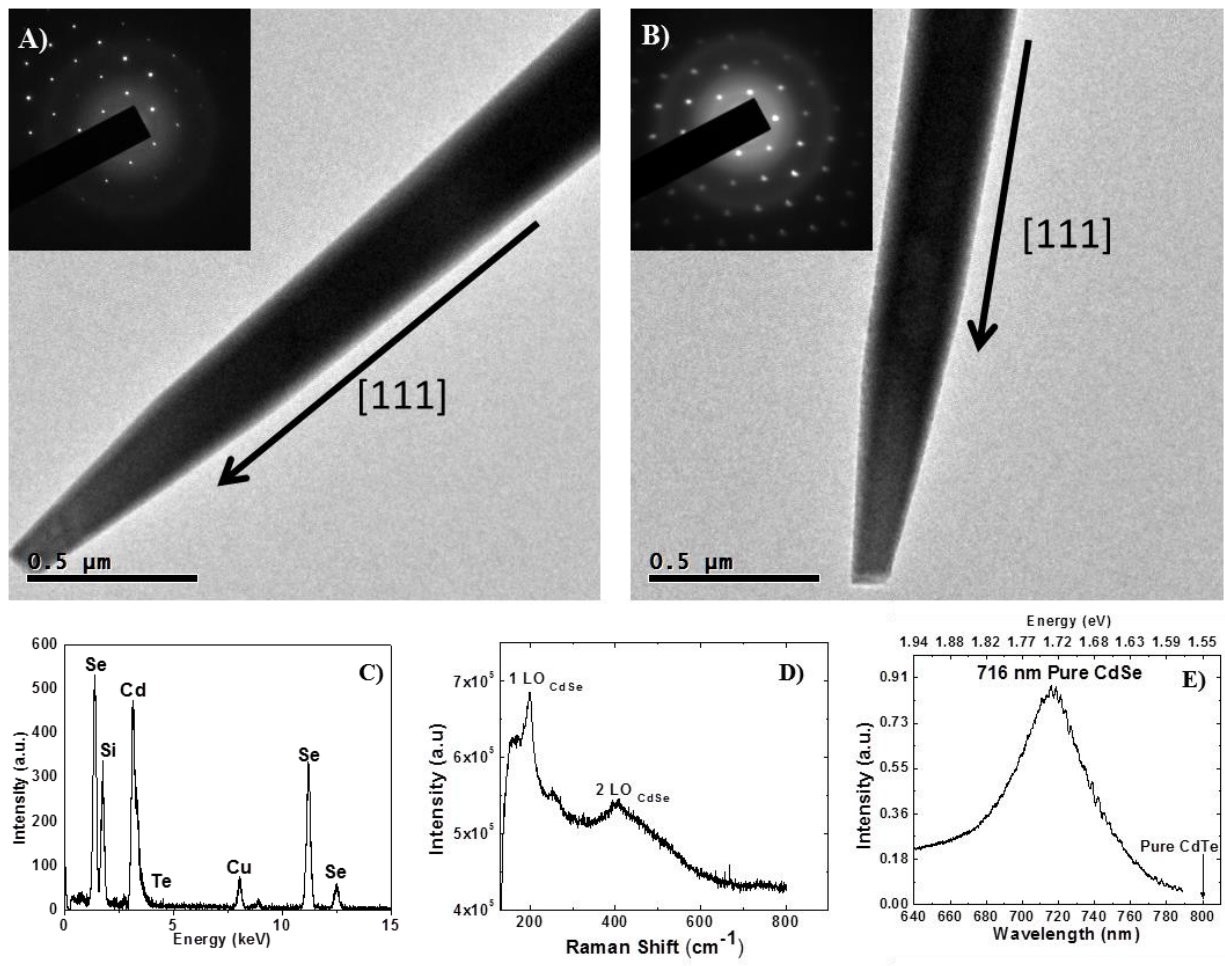

# Figure 3

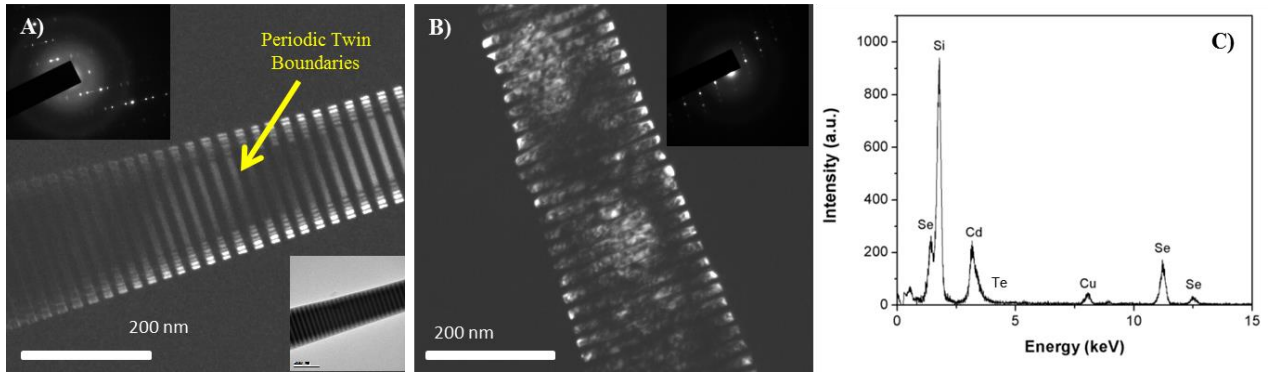

# Figure 4

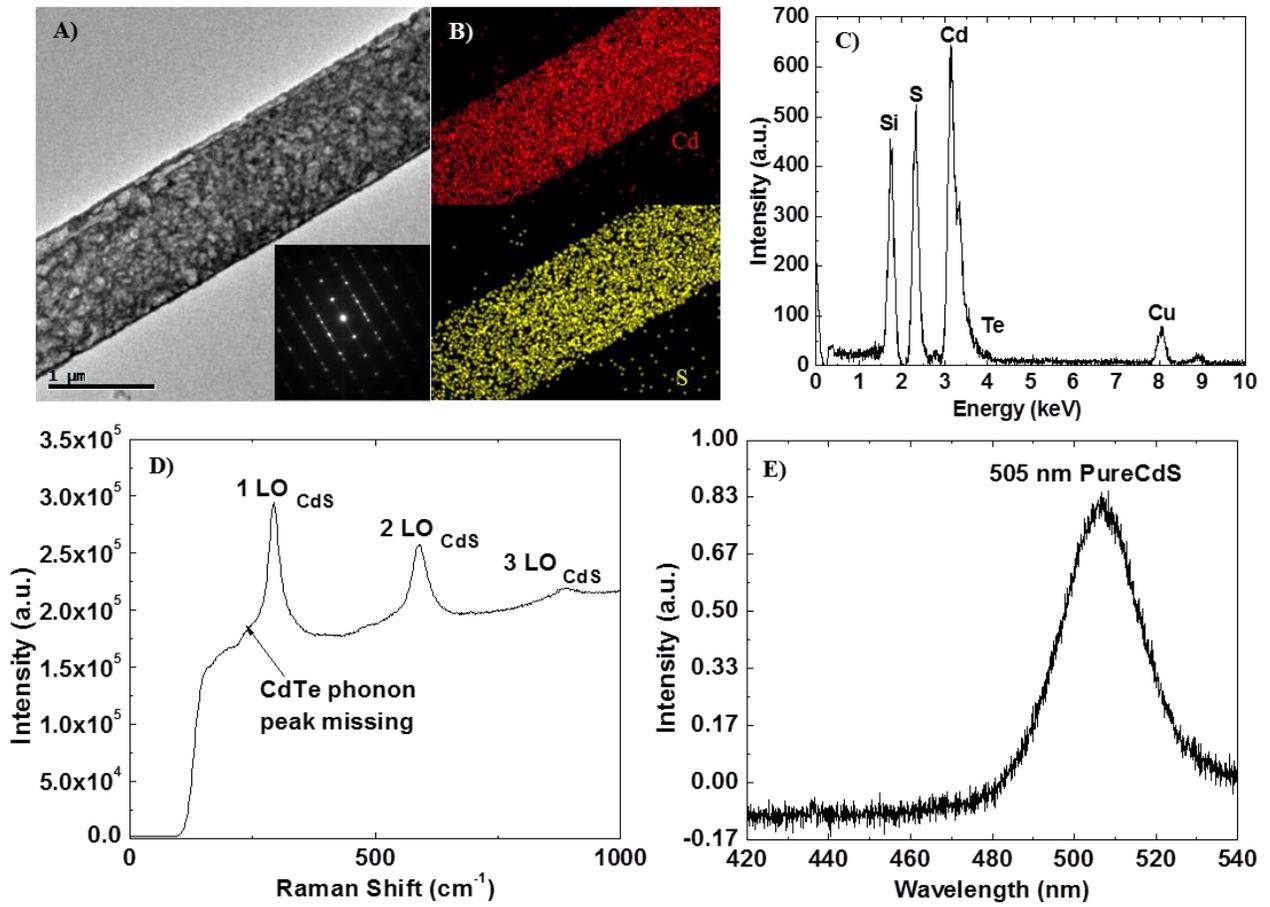

# Figure 5

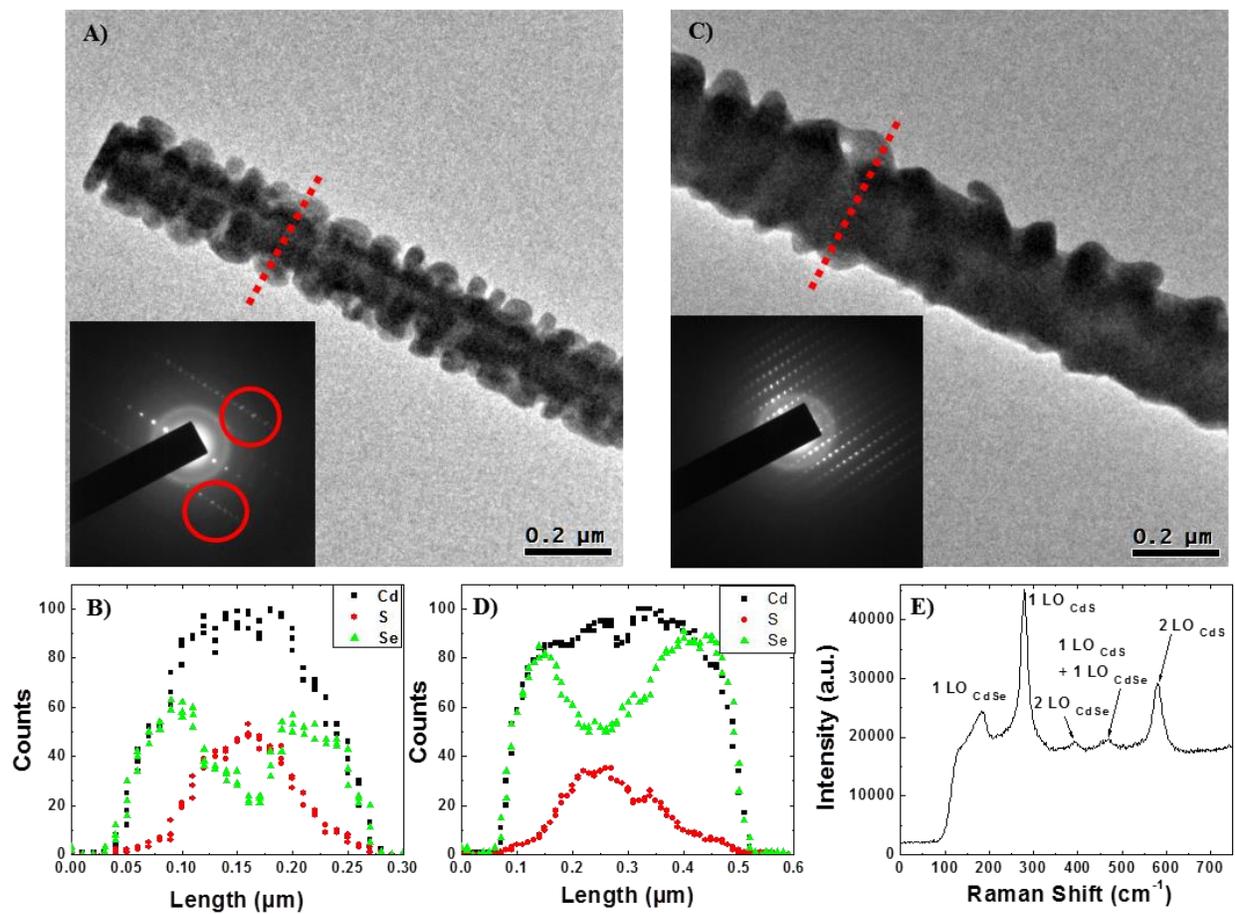

# Figure 6

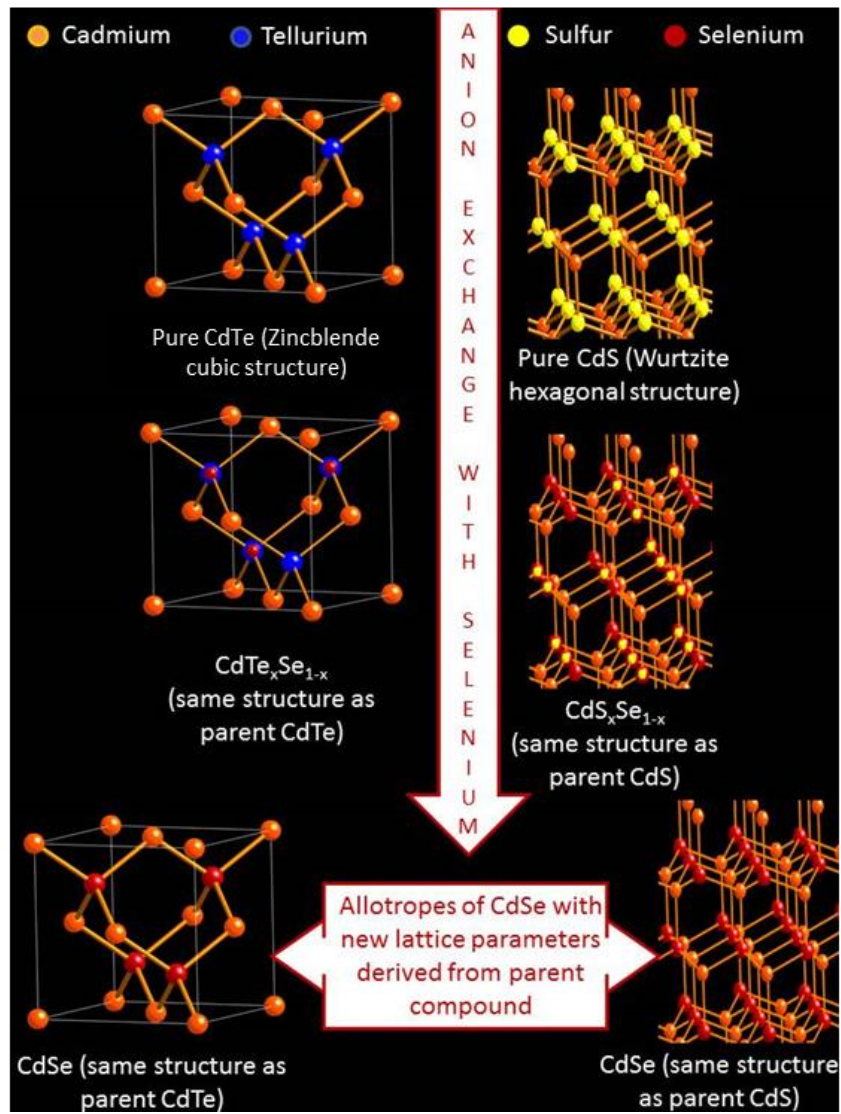

TOC Graphic

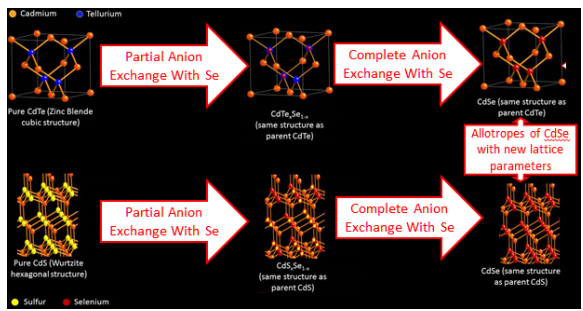

# Supplementary Information for

# Anion Exchange via Atomic Templating in II-VI Semiconducting Nanostructures

*Rahul Agarwal[1], Nadia M. Krook[1], Ming-Liang Ren[1], Liang Z. Tan[2], Wenjing Liu[1], Andrew M. Rappe[2], Ritesh Agarwal[1]* \*

[1]Department of Materials Science and Engineering, University of Pennsylvania, Philadelphia, PA 19104-6323, USA

[2] Department of Chemistry, University of Pennsylvania, Philadelphia, PA 19104-6323, USA

\* To whom correspondence should be addressed. E-mail: riteshag@seas.upenn.edu

**This file includes:**

TEM, Raman Spectroscopy and Photoluminescence experimental details

Growth and characterization details of CdTe

Anion exchange experimental details

Lattice constant calculation for ZB CdS and CdSe

Figures S1-S4

First-principles calculations using density functional theory

**1. Anion exchange experimental details**: As-grown nanostructures were dry transferred to a Si TEM grid with an electron beam transparent 50 nm thick SiN$_x$ window at the center. The grid was placed on the downstream side of the furnace such that the reaction temperature was uniform throughout the grid. 5 mg of the precursor (Se, S) in powdered form (Sigma-Aldrich) was loaded at the center of the furnace in a quartz tube and the furnace was pumped down to 20 mTorr. After flushing several times with Ar, the system was stabilized under a continuous flow of 15 SCCM and 5 Torr of Ar. The reaction temperature varied for different experiments but the reaction time was fixed at 1 hour (after the furnace reached the set temperature) after which the furnace was opened and left to cool until it reached ambient temperature. Reducing reaction time to 30 mins led to incomplete chemical transformation, whereas increasing reaction time to 90 mins resulted in substantial sublimation of the starting nanostructures for a given reaction temperature.

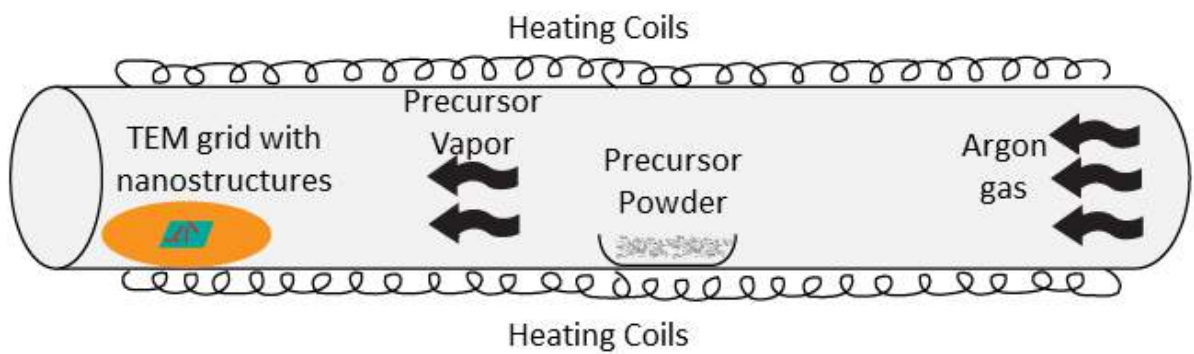

**Figure S1**: Schematic of the experimental set-up for anion exchange in a tube furnace.

**2. TEM, Raman Spectroscopy and Photoluminescence experiments**: Selected area electron diffraction (SAED) and HRTEM experiments were performed on JEOL 2100 TEM at an accelerating voltage of 200kV. Energy Dispersive Spectroscopy (EDS) study was performed on JEOL 2010F FEG TEM/STEM at an accelerating voltage of 200kV. Raman Spectroscopy (RS) was performed on a confocal micro-Raman system equipped with a 532 nm continuous-wave laser at ambient temperature and pressure. Photoluminescence experiments on individual nanostructures was carried out using a home-built optical microscopy set-up with a continuous-wave argon ion laser operating at 458 nm.

**3. Characterization of CdTe**:

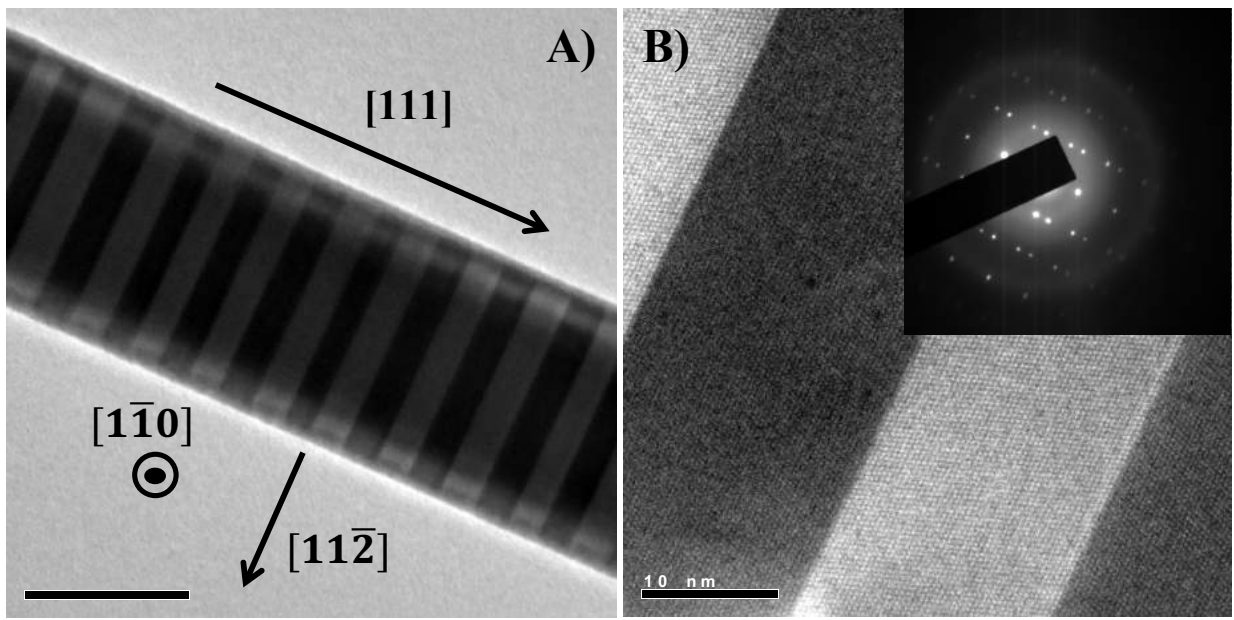

**Figure S2**: A) TEM micrograph of an as-grown CdTe nanowire grown along [111] with periodic twin boundaries running across the diameter throughout the length. B) HRTEM micrograph of the same nanowire showing phase contrast across twin boundaries in separate domains. Inset: SAED pattern of the same nanowire.

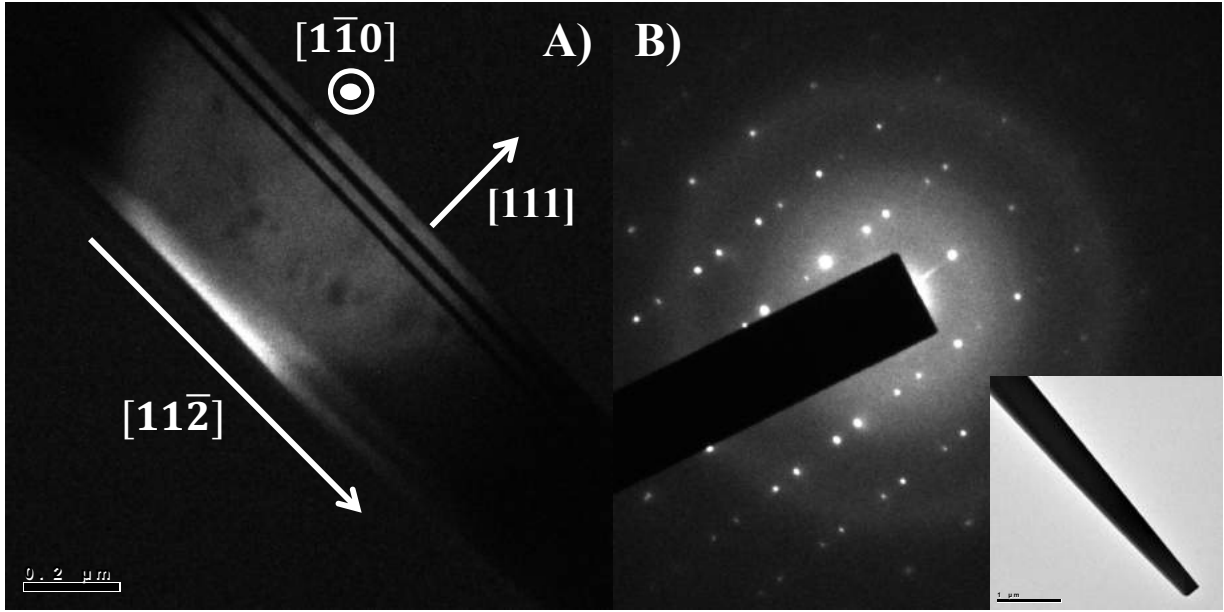

**Figure S3**: A) DF-TEM micrograph of an as-grown CdTe nanobelt grown along [112] with twin boundaries running across the length of the belt. B) SAED pattern of the same nanobelt confirming ZB crystal structure with twin boundaries present. Inset: BF-TEM micrograph of the nanobelt.

## 4. Lattice Constant Calculation for Zincblende CdS and CdSe

Let $a = ZB\ unit\ cell\ lattice\ constant$

$$a = \frac{4 * bond-distance}{\sqrt{3}}$$

$Cd - S \text{ bond distance for } WZ \text{ } CdS = 2.5 \text{ Å}$ [1]

$Cd - Se \text{ bond distance for } WZ \text{ } CdS = 2.6 \text{ Å}$ [2]

Therefore,

$$a_{CdS} = \frac{4 * 2.5}{\sqrt{3}} = 5.77 \text{ Å}$$

$$a_{CdSe} = \frac{4 * 2.6}{\sqrt{3}} = 6 \text{ Å}$$

## 5. Schematic of the mechanism of anion exchange

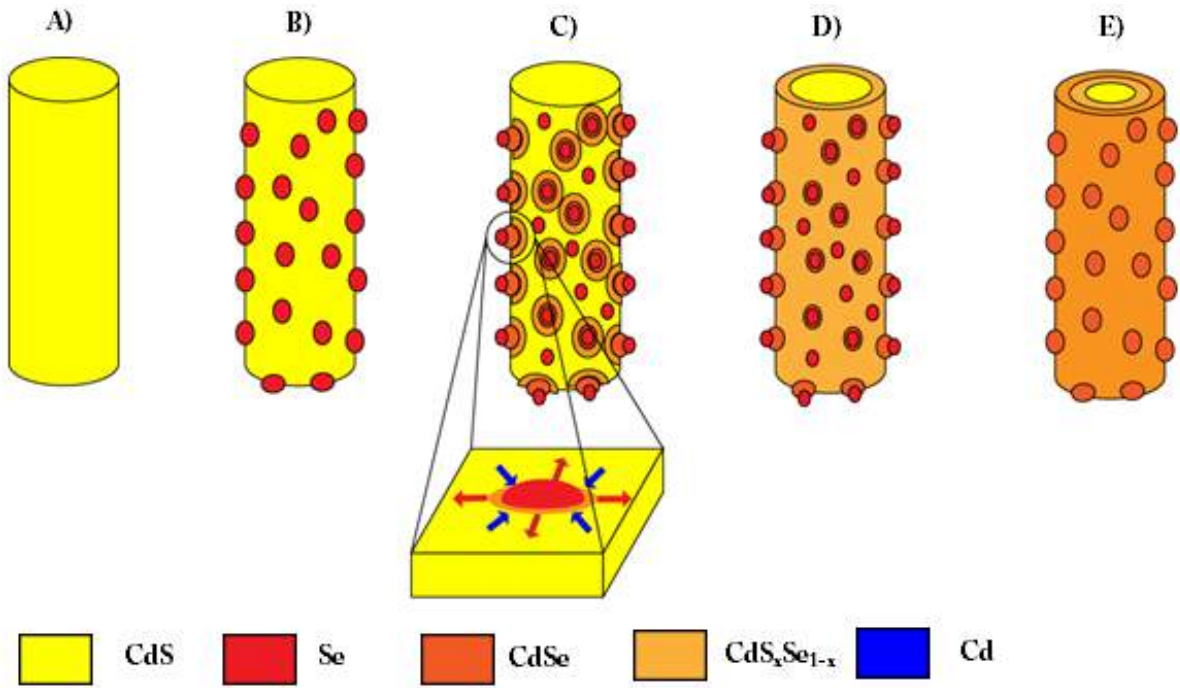

**Figure S4: Schematic of the proposed anion exchange mechanism.** A) Starting CdS nanostructure with smooth surfaces and well defined morphology. B) Condensation of the anion

precursor vapor (Se) on the surface of the nanostructure and beginning of anion exchange with surface ions ($S^{2-}$). C) Domain formation of $CdS_xSe_{1-x}$ alloy spreading through surface and bulk anion exchange. Inset: Zoomed-in schematic of surface diffusion process of various chemical species. D) Complete transformation of the surface into CdSe and the anion exchange reaction progressing towards the core. E) The reaction stops after either the precursor is exhausted or the system is cooled down.

## 6. First-principles calculations using density functional theory

We calculate the Gibbs free energy for the chemical reactions using first principles methods. For solid species, the Gibbs free energy is

$$G(\text{solid}) = E_e + F_{vib}$$

Where $E_e$ is the electronic contribution to the energy obtained from density functional theory (DFT) calculations and $F_{vib}$ is the vibrational contribution to the free energy [3], defined by

$$F_{vib} = \frac{1}{2}\sum_{nq} \hbar\omega_{nq} + k_B T \sum_{nq} \ln\left[1 - \exp(-\hbar\omega_{nq}/k_B T)\right]$$

Here, $\hbar\omega_{nq}$ are the phonon energies of branch $n$ and wavevector $q$. We treat monoatomic gaseous species as ideal gases, with their free energies given by

$$G(gas) = E_e - k_B T\left(1 + \ln\left[\frac{k_B T}{P}\left(\frac{2\pi m k_B T}{h^2}\right)^{3/2}\right]\right),$$

where $m$ is the mass of the atomic species and $P$ is the pressure.

Phonon frequencies were calculated within the density functional perturbation theory (DFPT) framework [4]. DFT and DFPT calculations were performed with the PBE exchange-correlation energy functional, using norm-conserving pseudopotentials. We used an 8x8x8 Monkhorst-Pack k-point grid to calculate the charge densities, and a 4x4x4 grid for the calculation of phonon frequencies. Calculations for the isolated atomic species were performed in a supercell geometry with a supercell size of 10 angstroms, while calculation of defect structures were calculated in a supercell geometry of 2x2x2 conventional unit cells.